\documentclass[twoside,showpacs,superscriptaddress,twocolumn,floatfix,a4paper,aps]{revtex4}

\usepackage{color}
\usepackage{graphicx}
\usepackage[utf8x]{inputenc}
\usepackage[T1]{fontenc}

\usepackage{amstext}
\usepackage{amsmath}            
\usepackage{amssymb}            

\newcommand{\etal}{\textit{et al.}}

\newcommand\extra[1]{}
\newcommand\red[1]{{#1}}

\begin{document}

\title{State-dependent linear-optical qubit amplifier}

\author{Karol Bartkiewicz} \email{bartkiewicz@jointlab.upol.cz}
\affiliation{RCPTM, Joint Laboratory of Optics of Palacký University and Institute of Physics of Academy of Sciences of the Czech Republic, 17. listopadu 12, 771 46 Olomouc, Czech Republic}

\author{Antonín Černoch}
\affiliation{Institute of Physics of Academy of Sciences of the Czech Republic, Joint Laboratory of Optics of PU and IP AS CR, 
   17. listopadu 50A, 772 07 Olomouc, Czech Republic}

\author{Karel Lemr}
\email{k.lemr@upol.cz}
\affiliation{RCPTM, Joint Laboratory of Optics of Palacký University and Institute of Physics of Academy of Sciences of the Czech Republic, 17. listopadu 12, 771 46 Olomouc, Czech Republic}

\date{\today}

\begin{abstract}
\red{We propose} a linear-optical setup for heralded qubit amplification with tunable output qubit fidelity. We study its success probability as a function of output qubit fidelity showing that at the expense of lower fidelity, the setup can \red{considerably} increase  probability of successful operation. These results are subsequently \red{applied in} a proposal for state dependent qubit amplification. Similarly to \red{state-dependent quantum cloning}, the \red{\textit{a priori}} information about the input state allows to optimize the \red{qubit amplification} procedure to obtain better fidelity versus success probability trade-off.
\end{abstract}


\maketitle

\section{Introduction}
Photons are \red{well suited to be} quantum information carriers \cite{gisin07comm}. Over the past decades, there has been a large number of both theoretically proposed and experimentally tested quantum information protocols \red{designed for photons} \cite{nielsen02info,braunstein05conti_info,bruss06info}. Notable example with practical applications is the quantum cryptography that allows for unconditionally secure transmission of information \cite{bb84,bb84exper,e91,ralph99CVcrypto,gisin02crypto}. \red{One can use both} both fiber \cite{wang12qkd} and free-space \cite{ursin07qkd} optics to distribute \red{photon-encoded} information over considerable distances. Even though \red{photons} are not so susceptible to interaction with the environment as for instance atoms \cite{specht11memory}, their state also deteriorates because of noise and absorption in the communication channel \cite{bartkiewicz07losses,halenkova12noise,horst13}.

Since channel transmissivity and level of noise are limited by unavoidable technological imperfections, a viable alternative strategy to increase communication range is based on amplification. \red{However,} quantum properties of photon states  \red{ (unless the state is known \textit{a priori}) are not preserved by classical amplification} based on mere ``measure and resend'' or stimulated emission approach\red{, thus these approaches are not always suitable \cite{oishi12ampl}}. Quantum amplifiers have to be used instead \cite{xiang10ampl,ferreyrol10ampl,zavata10ampl,osorio12ampl,micuda12ampl,mcmahon13ampl}.

In discrete variable encoding, polarization or spatial degree of freedom of individual photons are usually used to encode qubits. It is therefore not surprising that optical qubit amplifiers are proposed and built to address these degrees of freedom \cite{gisin10ampl,pitkanen11ampl,curty11ampl,kocsis13ampl,bula13qnd,scott13ampl}. Similarly to other linear-optical quantum gates \cite{brien07optical}, the qubit amplifiers are also probabilistic and their successful operation has to be heralded by specific detection outcome on ancillary photons. Thus apart from amplification gain, one has to introduce success probability to characterize performance of qubit amplifiers.

\begin{figure}
\includegraphics[width=6cm]{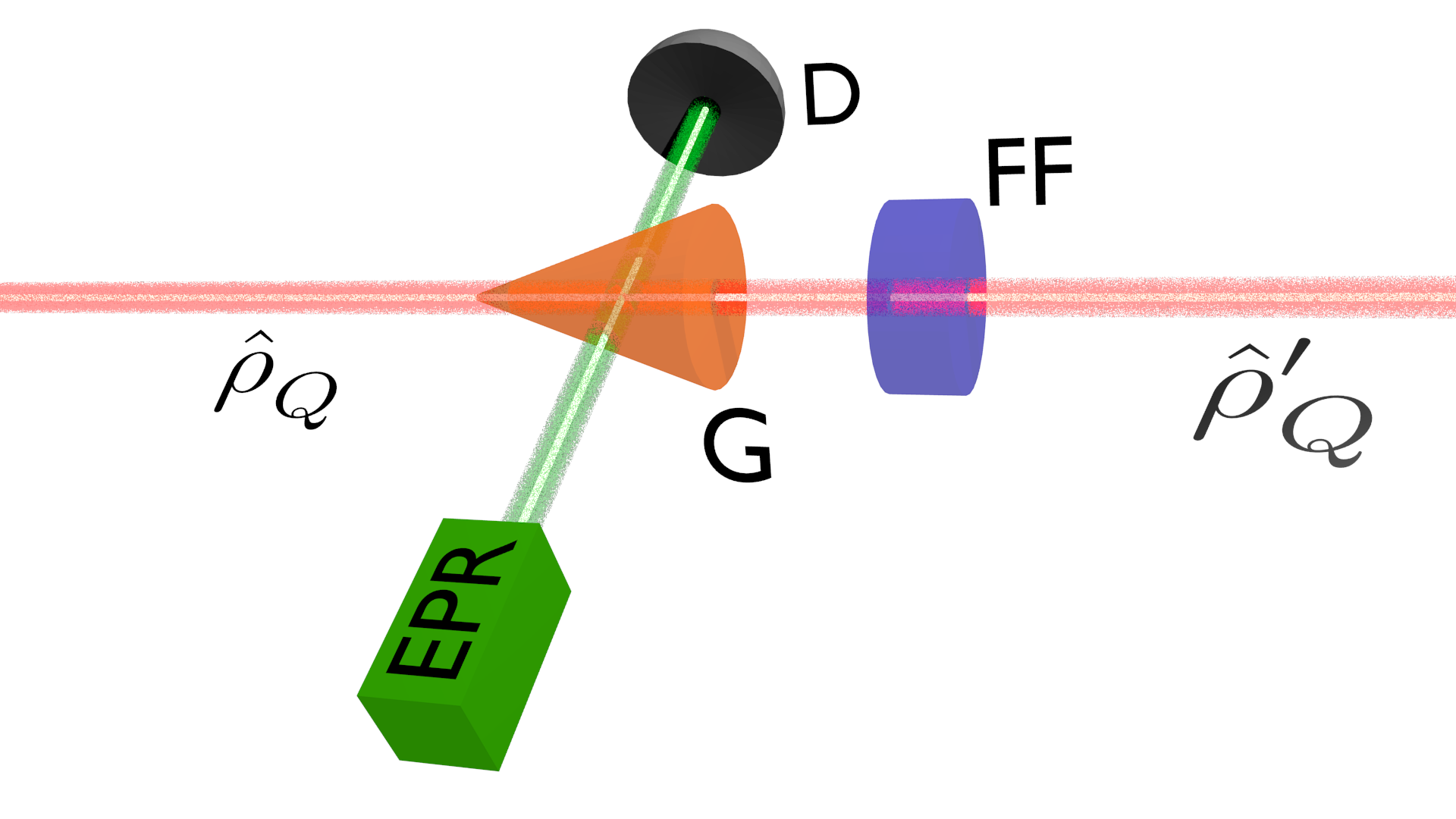}
\caption{\label{fig:concept} (\red{C}olor online) Conceptual scheme of a heralding qubit amplifier. Input state is transformed according to Eq. (\ref{eq:amplification}). D -- detector, EPR -- ancillary photons, G -- amplifier, FF -- feed forward.}
\end{figure}
In general, a qubit amplifier performs the following transformation on a mixture of vacuum and single qubit state
\begin{equation}
\label{eq:amplification}
p_0 |0\rangle\langle0| + p_1 \hat\rho_Q \rightarrow \frac{p_0}{N} |0\rangle\langle0| + \frac{p_1G}{N} \hat\rho'_Q,
\end{equation}
where $\hat\rho_Q$ and $\hat\rho'_Q$ stand for the input and output qubit density matrices, $N$ denotes normalization and $G$ is the overall \red{(nominal)} gain of the amplifier. So far only perfect amplifiers ($ \hat\rho_Q = \hat\rho'_Q$) have been discussed in literature. In this paper, we extend the analysis of our previously published scheme \cite{scott13ampl} to the general case of imperfect amplification ($ \hat\rho_Q \neq \hat\rho'_Q$).

\red{The paper is organized as follows: In
Sec.~\ref{sec_princip}, we describe the principle of operation of the proposed scheme. Moreover we introduce the describe the basic quantities used to characterize our proposed amplifier. We introduce fidelity of the operation as the overlap between the input and output qubit states. This analysis allows us to establish the success probability versus fidelity trade-off and observe increased success probability at the expense of a fidelity drop that we describe in Sec.~\ref{sec:trade-off}. Finally, in Sec.~\ref{sec:state-dependent}, inspired by \red{optimal} state-dependent quantum cloning \red{\cite{chiribella05cloner,bartkiewicz10cloner,lemr12cloner}}, we also show that having some \red{\textit{a priori}} information about the input state allows us to optimize the amplification procedure in order to improve this fidelity versus success probability trade-off. We conclude in Sec.~\ref{sec:conclusion}. }

\section{Principle of operation\label{sec_princip}}
\begin{figure}
\includegraphics[width=8.5cm]{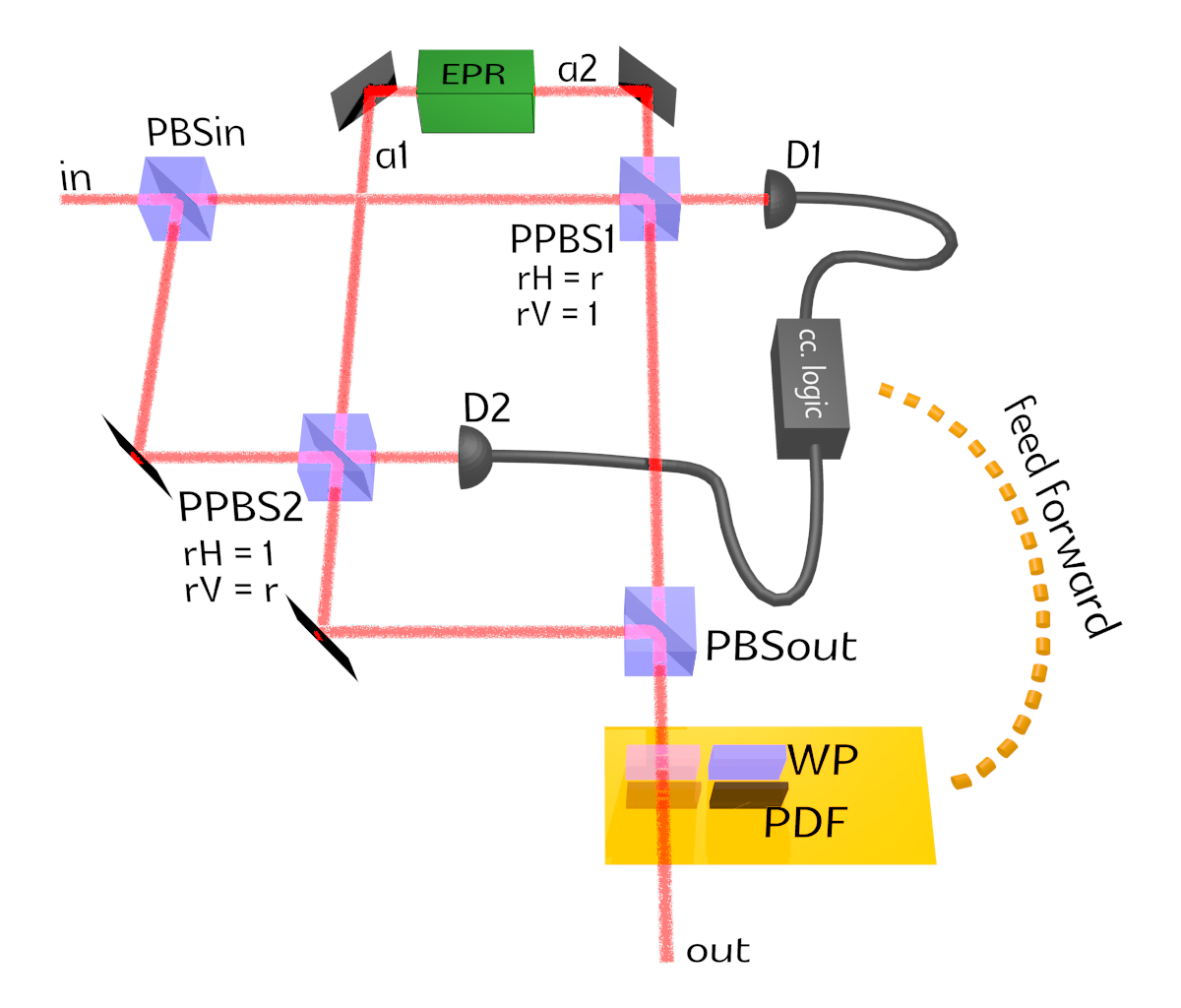}
\caption{\label{fig:scheme} (\red{C}olor online) Scheme for state-\red{dependent} linear-optical qubit amplifier as described in the text. EPR -- source of entangled ancillary photon pairs, PBS -- polarizing beam splitter, PPBS -- partially polarizing beam splitter (defined in the text), WP -- wave plate, PDF -- polarization dependent filter, D -- standard polarization analysis detection block (for reference see \cite{halenkova2012detector}).}
\end{figure}

In this section we describe the principle of operation of our scheme depicted in Fig. \ref{fig:scheme} so that in subsequent sections we can \red{analyze} the above mentioned fidelity vs. success probability trade-off and state dependent amplification.

The signal state $|\psi_s\rangle$ is prepared in superposition of vacuum $|0\rangle$ and single polarization encoded qubit state $|Q\rangle$
\begin{equation}
\label{eq:signal}
|\psi_s\rangle = \alpha |0\rangle +\beta |Q\rangle,
\end{equation}
where ($|\alpha|^2 + |\beta|^2 = 1$) and the qubit 
\begin{equation}
\label{eq:signalQ}
|Q\rangle = \cos\frac{\theta}{2}|H\rangle + \sin\frac{\theta}{2}\mathrm{e}^{i\varphi}|V\rangle
\end{equation}
\red{is parametrized by angles $\theta$ and $\varphi$ describing the superposition of horizontal $|H\rangle$  and vertical $|V\rangle$ polarization basis states.} The amplifier also makes use of an ancillary pair of entangled photons in a state parametrized by angle $\chi \in [0;\frac{\pi}{4}]$
\begin{equation}
\label{eq:ancilla}
|\psi_a\rangle = \cos\chi |HH\rangle + \sin\chi |VV\rangle.
\end{equation}

In the first step, the signal impinges on the first fully polarizing beam splitter \red{PBSin}, where the horizontal and vertical components of the signal qubit are separated into their respective modes. In these modes the interaction with the ancillary pairs of photons takes place: the horizontal component of the signal interacts with the first of the ancillary photons on a partially polarizing beam splitter \red{PPBS1}, similarly the vertical signal component is combined with the second ancillary photon on the partially polarizing beam splitter \red{PPBS2}. The partially polarizing beam splitter \red{PPBS1} fully reflects vertically polarized photons and has reflectivity $r$ for horizontal polarization. On the other hand, the \red{PPBS2} reflects all horizontally polarized light and has reflectivity $r$ for vertical polarization. Partially polarizing beam splitter \red{PPBS1} can be described in terms of creation operators
\begin{eqnarray*}
\hat{a}^\dagger_{\mathrm{in},H} & \rightarrow & r \hat{a}^\dagger_{\mathrm{out},H} + \sqrt{1-r^2} \hat{a}^\dagger_{D1,H} \\
\hat{a}^\dagger_{a1,H} & \rightarrow & -r \hat{a}^\dagger_{D1,H} + \sqrt{1-r^2} \hat{a}^\dagger_{\mathrm{out},H} \\
\hat{a}^\dagger_{a1,V} & \rightarrow & - \hat{a}^\dagger_{D1,V},
\end{eqnarray*}
where labelling of modes corresponds to the scheme in Fig. \ref{fig:scheme}. \red{Analogous transformation describes the action of the PPBS2.} Projection on diagonal \red{$|D\rangle = (|H\rangle + |V\rangle)/\sqrt{2}$} and anti-diagonal \red{$|A\rangle = (|H\rangle - |V\rangle)/\sqrt{2}$} linear polarization is performed in both detection modes $D1$ and $D2$. The resulting signal state is \red{recovered} by combing horizontal and vertical component on the output fully polarizing beam splitter \red{PBSout}.

One can trace how the individual components of the three-photon total state (signal and ancillary photons) get transformed by the setup assuming post-selection on detection of one photon in each detection mode $D1$ and $D2$
\begin{eqnarray*}
|0_\mathrm{in}H_{a1}H_{a2}\rangle &\rightarrow& r |0_\mathrm{out}H_{D1}H_{D2}\rangle \\
|0_\mathrm{in}V_{a1}V_{a2}\rangle &\rightarrow& r |0_\mathrm{out}V_{D1}V_{D2}\rangle \\
|H_\mathrm{in}H_{a1}H_{a2}\rangle &\rightarrow& (2r^2-1) |H_\mathrm{out}H_{D1}H_{D2}\rangle \\
|H_\mathrm{in}V_{a1}V_{a2}\rangle &\rightarrow& r^2 |H_\mathrm{out}V_{D1}V_{D2}\rangle \\
|V_\mathrm{in}H_{a1}H_{a2}\rangle &\rightarrow& r^2 |V_\mathrm{out}H_{D1}H_{D2}\rangle \\
|V_\mathrm{in}V_{a1}V_{a2}\rangle &\rightarrow& (2r^2-1) |V_\mathrm{out}V_{D1}V_{D2}\rangle .
\end{eqnarray*}

After \red{the photons} in the detection modes \red{get projected} to diagonal $|DD\rangle$ or anti-diagonal $|AA\rangle$ linear polarization states (both detected photons share the same polarization)
the output signal state \red{can be expressed as
\begin{eqnarray}
\label{eq:signalOUT1}
|\psi_\mathrm{out1}\rangle & \red{=} & \frac{\alpha r}{2} (\cos\chi + \sin\chi) |0\rangle \nonumber\\
&&+ \frac{\beta x_+}{2} \cos\frac{\theta}{2} |H\rangle + \frac{\beta y_+}{2} \sin\frac{\theta}{2} \mathrm{e}^{i\varphi} |V\rangle,
\end{eqnarray}
where 
\begin{eqnarray}
\label{eq:subst}
x_{\pm} &=& (2r^2-1) \cos{\chi} \pm r^2 \sin{\chi} \nonumber \\
y_{\pm} &=& (2r^2-1) \sin{\chi} \pm r^2 \cos{\chi}.
\end{eqnarray}
The output state $|\psi_\mathrm{out1}\rangle$ is kept intentionally not normalized to provide simple expression for success probability in subsequent calculations.} Alternatively the output signal state \red{(also not normalized)} takes the form of
\begin{eqnarray}
\label{eq:signalOUT2}
|\psi_\mathrm{out2}\rangle & \red{=} & \frac{\alpha r}{2} (\cos\chi - \sin\chi) |0\rangle  \nonumber\\
&&+ \frac{\beta x_-}{2} \cos\frac{\theta}{2} |H\rangle - \frac{\beta y_-}{2} \sin\frac{\theta}{2} \mathrm{e}^{i\varphi} |V\rangle
\end{eqnarray}
if $|DA\rangle$ or $|AD\rangle$ coincidence is observed (detected photons have mutually orthogonal polarizations).

A feed-forward operation \red{has} to be adopted to correct the qubit part of state \red{given by Eq.~(\ref{eq:signalOUT1})} to be identical to qubit part of \red{Eq.~(\ref{eq:signalOUT2})}. This feed-forward transformation consists of polarization dependent filtrations $\tau_H$ and $\tau_V$ when $|DD\rangle$ or $|AA\rangle$ coincidence are detected. These filtrations are functions of the ancilla parameter $\chi$ and reflectivity $r$, but are signal state independent:
\begin{equation}
\label{eq:filtration}
\tau_H = \frac{x_-}{x_+}, \;\;\; \tau_V = \frac{y_-}{y_+} .
\end{equation}
\red{In} the case of $|DA\rangle$ or $|AD\rangle$ coincidence detection, additional phase shift (sign flip) is imposed to vertical polarization ($V \rightarrow -V$). This process is not lossy so we assume it is performed in all the subsequently evaluated scenarios.

\red{\subsection*{Success probability}}

For the subsequent analysis, several quantities are crucial. First of them is the overall success probability of the procedure $P_\mathrm{succ}$. It can be expressed using the norm of the output state $|\psi_\mathrm{out1}\rangle$ and $|\psi_\mathrm{out2}\rangle$. Not implementing the lossy feed-forward, the success probability reads
\begin{eqnarray}
\label{eq:psucc}
P_\mathrm{succ} &=& 2(|\langle\psi_\mathrm{out1}|\psi_\mathrm{out1}\rangle|+|\langle\psi_\mathrm{out2}|\psi_\mathrm{out2}\rangle|) \nonumber \\
&=& |\alpha|^2 r^2 + |\beta|^2 \frac{x_+^2 + x_-^2}{2} \cos^2{\frac{\theta}{2}}  \nonumber \\
&&+ |\beta|^2 \frac{y_+^2 + y_-^2}{2} \sin^2{\frac{\theta}{2}} ,
\end{eqnarray}
\red{where the factor of two describes the two equally probable coincidences leading to $|\psi_\mathrm{out1}\rangle$ or $|\psi_\mathrm{out1}\rangle$.} On the other hand, if the feed-forward is implemented, the output states $|\psi_\mathrm{out1}\rangle$ and $|\psi_\mathrm{out2}\rangle$ are transformed to the form of
\begin{eqnarray}
\label{eq:signalOUT2FF}
|\psi_\mathrm{out1FF}\rangle & \red{=} & \frac{\alpha r}{2} \left(\cos\chi + \sin\chi\right) |0\rangle  \nonumber\\
&& + \frac{\beta x_-}{2} \cos\frac{\theta}{2} |H\rangle + \frac{\beta y_-}{2} \sin\frac{\theta}{2} |V\rangle , \nonumber \\
|\psi_\mathrm{out2FF}\rangle & \red{=} & \frac{\alpha r}{2} \left(\cos\chi - \sin\chi\right) |0\rangle + \nonumber\\
&&+ \frac{\beta x_-}{2} \cos\frac{\theta}{2} |H\rangle + \frac{\beta y_-}{2} \sin\frac{\theta}{2} |V\rangle 
\end{eqnarray}
and the corresponding success probability reads
\begin{eqnarray}
\label{eq:psuccFF}
P_\mathrm{succ} &=& 2(|\langle\psi_\mathrm{out1FF}|\psi_\mathrm{out1FF}\rangle|+|\langle\psi_\mathrm{out2FF}|\psi_\mathrm{out2FF}\rangle|)  \nonumber \\
&=& |\alpha|^2 r^2 + |\beta|^2 \left( x_-^2 \cos^2{\frac{\theta}{2}} + y_-^2 \sin^2{\frac{\theta}{2}} \right).
\end{eqnarray}

\red{\subsection*{Amplification gain}}

A second very important parameter of the amplifier is the gain \red{$G$} -- \red{the ratio between qubit and vacuum components for the amplified state divided by the analogous ratio for the} initial input state \red{as show in Eq.~(\ref{eq:amplification})}. In general, the gain can differ for horizontal and vertical polarizations. One can easily define the gain for both polarizations in the case the feed-forward is implemented
\begin{equation}
\label{eq:gainFF}
G_{HFF} = \frac{x_-^2}{r^2}, \;\;\;
G_{VFF} = \frac{y_-^2}{r^2}.
\end{equation}
If the lossy feed-forward is not implemented, the gain can be calculated as average gain for output state $|\psi_\mathrm{out1}\rangle$ and $|\psi_\mathrm{out2}\rangle$
\begin{equation}
\label{eq:gain}
G_{H} = \frac{x_+^2+x_-^2}{2r^2}, \;\;\;
G_{V} = \frac{y_+^2+y_-^2}{2r^2}.
\end{equation}
\red{The} overall gain defined in Eq.~(\ref{eq:amplification}) \red{is obtained by combining} the two gains for horizontal and vertical polarization. In the case of applied feed-forward, the overall gain is given by
$$
G_{FF} = \cos^2\frac{\theta}{2}G_{HFF} + \sin^2\frac{\theta}{2}G_{VFF}
$$
and in the case without the lossy feed-forward \red{(only the phase flip performed)} it is given similarly by
$$
G = \cos^2\frac{\theta}{2}G_{H} + \sin^2\frac{\theta}{2}G_{V}.
$$

\red{\subsection*{Amplification fidelity}}

The \red{last} quantity that has to be calculated in this section is the output qubit fidelity $F_Q$. This fidelity compares the overlap between the qubit state $|Q\rangle$ at the input with the qubit subspace of the output state $|\psi_\mathrm{outQ}\rangle$. If the feed-forward is implemented, the fidelity is simply
\begin{equation}
\label{eq:fidelityFF}
F_\mathrm{QFF} = |\langle\psi_\mathrm{outQ}|Q\rangle|^2 = \frac{\left( x_- \cos^2{\frac{\theta}{2}} + y_- \sin^2{\frac{\theta}{2}} \right)^2}{x_-^2 \cos^2{\frac{\theta}{2}} + y_-^2 \sin^2{\frac{\theta}{2}}}.
\end{equation}
If only the feed-forward phase correction and not the \red{full} lossy transformation is performed, the fidelity of the output qubit reads
\red{
\begin{eqnarray}
\label{eq:fidelity}
\!\!\!\!&&F_\mathrm{Q}= \langle Q|\hat\rho_\mathrm{outQ}|Q\rangle \\ \nonumber
\!\!\!\!&&= \frac{\left( x_+ \cos^2{\frac{\theta}{2}} + y_+ \sin^2{\frac{\theta}{2}} \right)^2 + \left( x_- \cos^2{\frac{\theta}{2}} + y_- \sin^2{\frac{\theta}{2}} \right)^2}{(x_+^2 + x_-^2) \cos^2{\frac{\theta}{2}} + (y_+^2 + y_-^2) \sin^2{\frac{\theta}{2}}},
\end{eqnarray}} 
where $\hat\rho_\mathrm{outQ}$ is the \red{normalized} density matrix \red{of the single photon subspace being a balanced} mixture of $|\psi_\mathrm{out1}\rangle\langle\psi_\mathrm{out1}|$ and  $|\psi_\mathrm{out2}\rangle\langle\psi_\mathrm{out2}|$ with $V\rightarrow-V$ transformation performed on the later.

\section{Success probability--fidelity trade-off \label{sec:trade-off}}
\begin{figure}[t]
\includegraphics[scale=1]{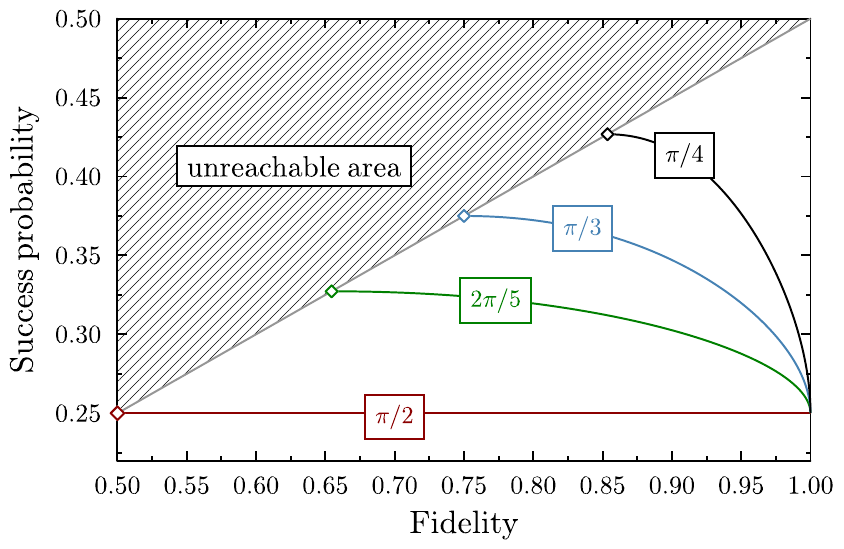}
\caption{\label{fig:psucc_fid_Infgain} (\red{C}olor online) Success probability \red{$P_\mathrm{succ}$ given by Eq.~(\ref{eq:r0Psucc})} as a function of output state fidelity \red{$F_\mathrm{QFF}$ given by Eq.~(\ref{eq:r0fidelity})} in the case of infinite gain is depicted for four different input states as described in the text.}
\end{figure}
\begin{figure}[t]
\includegraphics[scale=1]{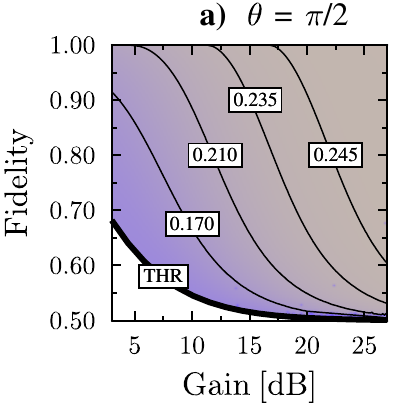}
\includegraphics[scale=1]{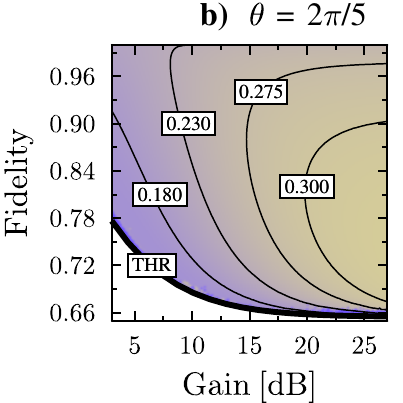}\\\vspace{1em}
\includegraphics[scale=1]{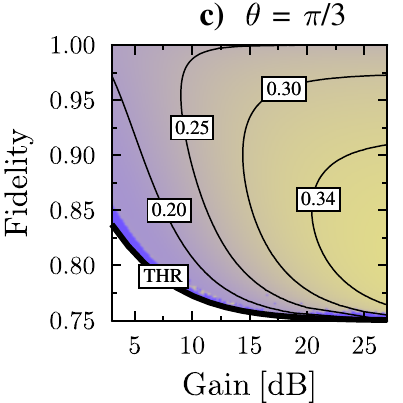}
\includegraphics[scale=1]{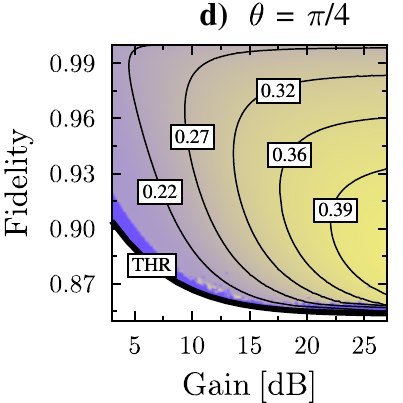}
\caption{\label{fig:psucc_fid_gain} (\red{C}olor online) Success probability \red{$P_{\mathrm{succ}}$} as a function of both output state fidelity and amplification gain \red{$G_{\mathrm{FF}}$} is depicted for four different input states as described in the text. THR stands for threshold of unreachable area.}
\end{figure}

In this section we investigate the trade-off between success probability \red{$P_{\mathrm{succ}}$} and the output state fidelity \red{$F_{\mathrm{QFF}}$}. For this analysis, we fixed the parameters $\alpha = \beta = \frac{1}{\sqrt{2}}$ and we also took into account the lossy feed-forward. 

\subsection*{Infinite gain}

\red{First}, we studied this trade-off on the particular case of infinite gain. The infinite gain is \red{an important} setting of qubit amplifiers. To achieve \red{this} regime, one simply sets $r = 0$. \red{Thus, the } previously obtained \red{expressions} can be considerably simplified. Coefficients $x_+ = x_- = - \cos{\chi}$ and $y_+ = y_- = - \sin{\chi}$ become equal so there is no need for lossy feed-forward any more ($\tau_H = \tau_V = 1$), only the $V\rightarrow-V$ is performed. Success probability and qubit fidelity take the form of
\begin{eqnarray}
\label{eq:r0Psucc}
P_\mathrm{succ}&=&|\beta|^2 \left( \cos^2{\chi} \cos^2{\frac{\theta}{2}} + \sin^2{\chi} \sin^2{\frac{\theta}{2}} \right)  \nonumber \\
&=& \frac{|\beta|^2}{2} \left[ \cos^2{\left( \chi - \frac{\theta}{2} \right)} + \cos^2{\left( \chi + \frac{\theta}{2} \right)} \right]
\end{eqnarray}
and
\begin{equation}
\label{eq:r0fidelity}
F_\mathrm{QFF} = \frac{\left( \cos{\chi} \cos^2{\frac{\theta}{2}} + \sin{\chi} \sin^2{\frac{\theta}{2}} \right)^2}{\cos^2{\chi} \cos^2{\frac{\theta}{2}} + \sin^2{\chi} \sin^2{\frac{\theta}{2}}}
\end{equation}
respectively. 

The Fig.~\ref{fig:psucc_fid_Infgain} shows the dependence of the success probability on output state fidelity for four different input state parametrized by $\theta = \lbrace \pi/2, 2\pi/5, \pi/3, \pi/4 \rbrace$ and $\varphi = 0$. Calculation reveals that there is no improvement in success probability in the case of a balanced input state ($\theta = \pi/2$) and the success probability remains constant and fidelity independent. In contrast to that, the more the input state is unbalanced, the more pronounced is the dependence of the success probability on fidelity. This fact will \red{reemerge} in the next section discussing state dependent amplification. For instance in the case of $\theta = \pi/4$, the success probability can be increased by a factor of 1.7 at the expense of 85\% output state fidelity.

\red{\subsection*{Maximum success probability}}

In the next step, we performed numerical calculation of maximum achievable success probability for given values of overall gain \red{given by Eq.~(\ref{eq:amplification})} and \red{the} output state fidelity \red{given by Eq.~(\ref{eq:fidelityFF})}. This calculation has been carried out on the same four input state as mentioned above \red{by varying the $\chi$ and $r$ parameters}. Plots in Fig.~\ref{fig:psucc_fid_gain} present the obtained results confirming the finding described in Fig. \ref{fig:psucc_fid_Infgain}. In addition to that, one can observe that set to lower values of gain, the setup performs better for higher fidelities than for lower ones. In the case of higher gains however, the setup behaves as described in the infinite gain analysis. Also we were able to establish \red{state-dependent} unreachable area -- set of gain and fidelity coordinates that can not be reached by presented setup. This area is visualized by the threshold \red{(TRH) line shown in Fig.~\ref{fig:psucc_fid_gain}}.

\section{State-dependent amplification \label{sec:state-dependent}}
\begin{figure}[t]
\includegraphics[scale=0.9]{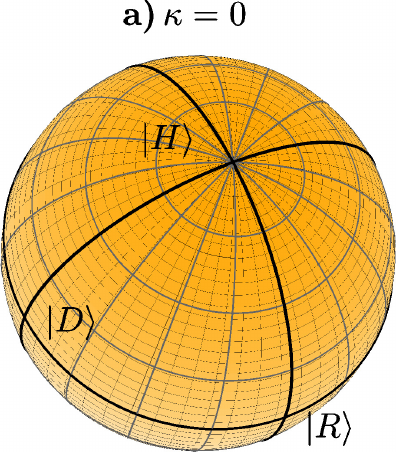}\hspace{1em}
\includegraphics[scale=0.9]{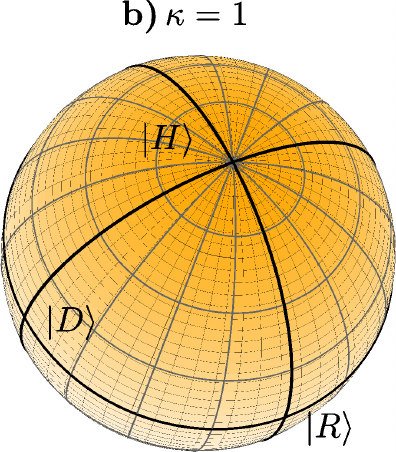}\\\vspace{2em}
\includegraphics[scale=0.9]{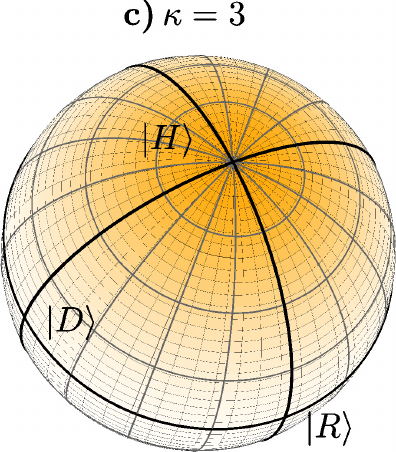}\hspace{1em}
\includegraphics[scale=0.9]{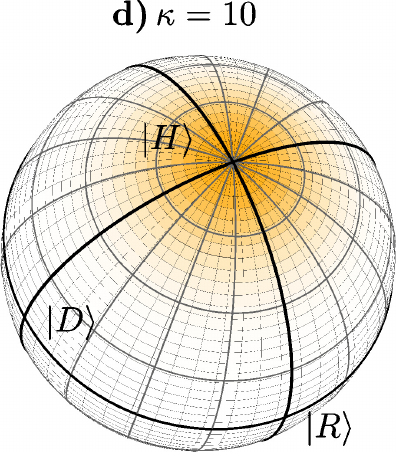}
\caption{\label{fig:spheres} (\red{C}olor online) Probability density function $g = g(\theta,\kappa)$ \red{given by Eq.~(\ref{eq:density})} over the Poincaré sphere for various values of $\kappa$ parameter used in subsequent numerical simulations. Labels $|H\rangle$, $|D\rangle$ and \red{$|R\rangle=(|H\rangle +i|V\rangle)/2$} denote position of horizontal, diagonal and right-hand circular polarization states respectively.}
\end{figure}
\begin{figure}
\includegraphics[scale=1]{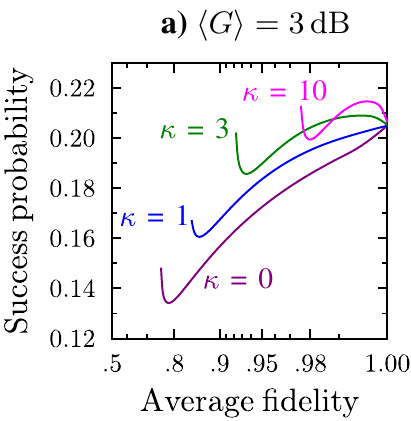}
\includegraphics[scale=1]{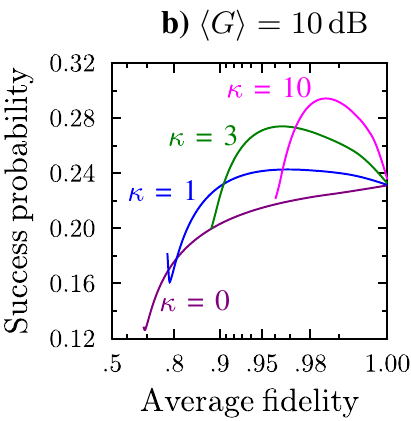}\\\vspace{1em}
\includegraphics[scale=1]{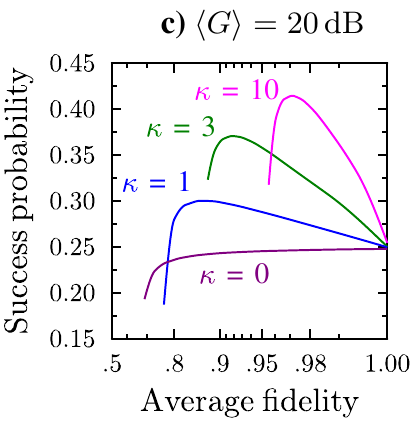}
\includegraphics[scale=1]{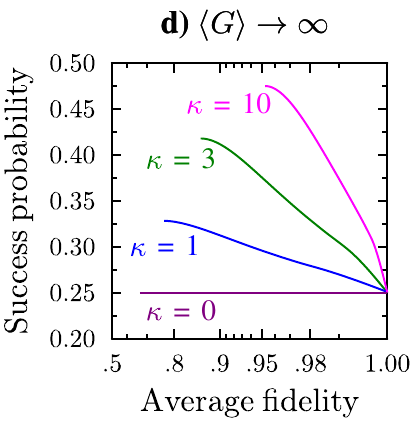}
\caption{\label{fig:psucc_fid_kappa} (\red{C}olor online) \red{Maximum achievable success probability $\langle P\rangle$} as a function of average fidelity \red{$\langle F\rangle$} for various values of average overall gain $\langle G\rangle$ and state knowledge described by parameter $\kappa$ of probability density function $g = g(\theta,\kappa)$ \red{given by Eq.~(\ref{eq:density})}.}
\end{figure}

This section brings forward the main result of the paper: how can we improve the success probability of amplification given some \red{\textit{a priori}} knowledge about the input qubit state? For the purpose of quantifying the \red{\textit{a priori}} information about the input signal, we use the \red{Von Mises–Fisher} distribution \red{\cite{Fisher} (also known as the Kent distribution) describing dispersion on a sphere. This} probability density function is defined \red{as}
\begin{equation}
\label{eq:density}
g(\theta,\kappa) = \frac{\kappa}{4\pi\,\mathrm{sinh}(\kappa)}\mathrm{exp}(\kappa\cos\theta),
\end{equation}
where $\theta$ is the input state parameter \red{describing the axial angle of the state on the Poincaré sphere} and $\kappa$\red{, i.e. the concentration parameter,} determines the amount of knowledge about the input qubit. The Fig. \ref{fig:spheres} depicts the probability distribution over the Poincaré sphere for various values of $\kappa$. Note that in the case of $\kappa = 0$, all states are equally probable (therefore no \textit{a priori} knowledge) and
the larger the \red{concentration parameter} $\kappa$ is, the more precise information about the input state we have. \red{This trend is illustrated in} the \red{Tab.~\ref{tab:kent} providing the} values of medians $\theta_m$ and first deciles $\theta_d$ for various values of $\kappa$. Note that while throughout this paper we center the distribution $g(\kappa,\theta)$ around the \red{northern} pole of the sphere -- horizontal polarization -- the generality of our scheme does not suffer by this choice. If \red{the} knowledge about the input state is not centred around north pole, one can always perform a deterministic rotation to make it so and inverse it after the state comes out of the amplifier.
\begin{table}
\caption{\label{tab:kent}Values of medians and first deciles of the \red{the Von Mises-Fisher} distribution for several values of \red{the concentration} parameter $\kappa$.}
\begin{ruledtabular}
\begin{tabular}{lll}
Parameter $\kappa$ & Median $\theta_m$ [rad] & First decile $\theta_d$ [rad]\\
0 & $\pi/2$ & $0.205\pi$\\
1 & $0.357\pi$ & $0.136\pi$\\
3 & $0.220\pi$ & $0.085\pi$\\
10 & $0.119\pi$ & $0.046\pi$
\end{tabular}
\end{ruledtabular}
\end{table}
Using this quantification of input state knowledge, we performed a series of numerical \red{calculations} with the goal to determine the fidelity--success probability trade-offs. \red{Our results show the relation between the highest achievable average success probability
\begin{eqnarray*}
\langle P_{\mathrm{succ}} \rangle = \int_\Omega g(\theta,\kappa) P_{\mathrm{succ}}\,\mathrm{d}\omega,  
\end{eqnarray*}
for the fixed values of average gain and fidelity
\begin{eqnarray*}
\langle G \rangle &=& \int_\Omega g(\theta,\kappa) G_{\mathrm{FF}}\,\mathrm{d}\omega, \\
\langle F \rangle &=& \int_\Omega g(\theta,\kappa) F_{\mathrm{QFF}}\,\mathrm{d}\omega,  
\end{eqnarray*}
respectively, where $\mathrm{d}\omega=-\mathrm{d}\cos\theta\,\mathrm{d}\phi$ and $\Omega$ is the surface of the Poincaré sphere. Only the $\langle F \rangle$ integral is not trivial since $F_{\mathrm{QFF}}$ is a rational function of $\cos\theta$, thus it was calculated numerically, however the other integrals can be expressed as linear functions of $\langle\cos\theta\rangle = \coth\kappa - 1/\kappa$.  The investigated cases} are depicted in Fig~\ref{fig:psucc_fid_kappa}. In each \red{case} we targeted one specific average \red{overall} gain \red{value from the set $\langle G\rangle \in \lbrace 3\,\mathrm{dB},10\,\mathrm{dB},20\,\mathrm{dB},\infty\rbrace$, where the average was taken over input states distributed according to the Von Mises-Fisher distribution for} four different values of $\kappa \in\lbrace0;1;3;10\rbrace$. For all the average gain and $\kappa$ combinations, we determined the relation between \red{the} average output state fidelity and \red{the average} success probability. Note that similarly \red{as in} the previous section, we assumed $\alpha = \beta = \frac{1}{\sqrt{2}}$ and we also took into account the lossy feed-forward.

Similarly \red{as in the case analyzed in Sec.~\ref{sec:trade-off}}, not all 
\red{the} values of fidelity are accessible simply because of the fact, that 
the setup can not produce fidelity lower that a certain threshold that 
depends on the values of $\kappa$ and average gain. It is a well expected 
result, that for the combination of $\kappa = 0$ and infinite gain, \red{the 
success probability of the setup and fidelity are state-independent}. This 
result can be analytically verified using formulas from 
\red{Sec.~\ref{sec_princip} for} $r = 0$. In contrast to that, for other 
\red{than} infinite average gains, there is always a maximum of success 
probability depending on $\kappa$. For $\kappa\rightarrow\infty$, this 
maximum is found for unit fidelity \red{$\langle F\rangle = 1$}. It follows 
from the above mentioned observations that for a given value of average gain 
\red{$\langle G\rangle$} and $\kappa$, there \red{exists} a specific fidelity 
\red{value $\langle F \rangle$} giving maximum success probability 
\red{$\max_{\langle F \rangle}{\langle P\rangle}=P_{\mathrm{max}}$}. In some cases this maximum \red{is 
to be found on the threshold providing the lower bound on the accessible fidelity values, but surprisingly this is not always the case}. This effect 
reflects the fact that \red{the space of $\chi$ and $r$ values providing at the same time the required value of fidelity and the average gain has a non-trivial structure. Thus, it seams that the question about the the limits on the success rate of the state-dependent quantum amplifier for fixed amplification parameters does not have a simple answer. Nevertheless, it is apparent that in general one can increase the success probability of the setup at the expense of the lower success probability, but sometimes the maximum value $P_{\mathrm{max}}$ can be reached at a lower cost than approaching the fidelity threshold.} 

\red{\subsection*{Merit function}}

\begin{figure}[t]
\includegraphics[scale=1]{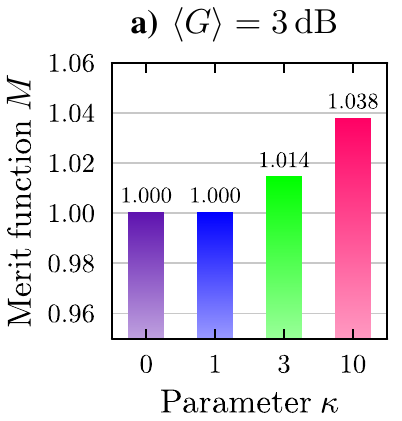}
\includegraphics[scale=1]{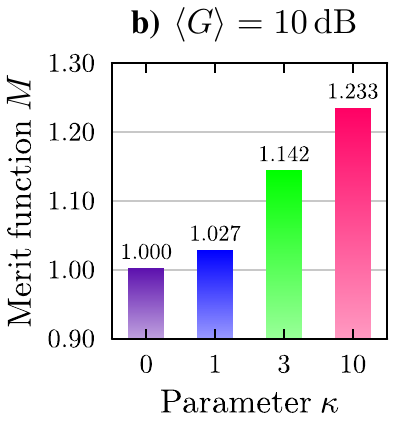}\\\vspace{1em}
\includegraphics[scale=1]{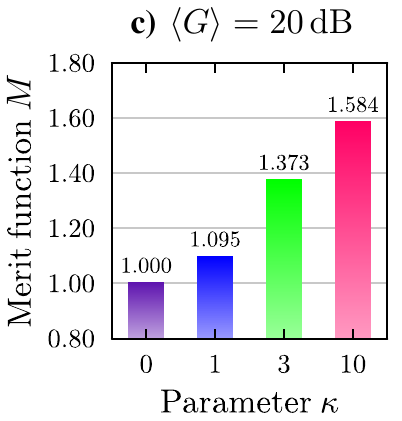}
\includegraphics[scale=1]{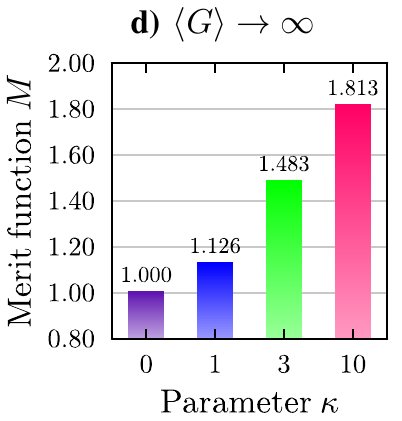}
\caption{\label{fig:kappa_merit} (\red{C}olor online) Merit function $M$ \red{given by Eq.~(\ref{eq:merit})} depicted for various parameters $\kappa$ and average gains $\langle G\rangle$.}
\end{figure}
One can argue, that some applications require perfect amplification with unit fidelity and thus it is not suitable to increase the success probability of the setup at the expense of lower fidelity. While this may indeed be true in some cases, realistic protocols for quantum communication have to be robust against at least some degree of fidelity drop. This leads us to formulate a figure of merit function inspired by \cite{lemr12eff}
\begin{equation}
M = \frac{\mathrm{max} \lbrace P_\mathrm{succ}F\rbrace}{P_\mathrm{succ}({F=1})},\label{eq:merit}
\end{equation}
where the numerator is the maximum of the product of fidelity and corresponding success probability and the denominator is just the success probability at unit fidelity. Since the product of fidelity and success probability can be understood as some sort of output rate of signal qubits, the function $M$ gives maximum factor of increased output signal rate if one allows for the fidelity to be smaller then 1 (see Fig.~\ref{fig:kappa_merit}).

\red{It can be easily shown that} for the very specific case of both infinite average gain and infinite $\kappa$, the setup gives exactly the same outcomes of simple {\em photon amplifier} \cite{ralph09ampl} based on the ``detect and reproduce'' method. While for no \red{\textit{a priori}} knowledge about the input state $\kappa = 0$, the setup provides the same \red{functionality} as previously published {\em qubit amplifier} \cite{scott13ampl}. In this sense, the setup covers the transition between these two conceptually different devices.

\section{Conclusions \label{sec:conclusion}}
The possibility to operate a qubit amplifier in a imperfect regime, where output qubit fidelity may be smaller than one offers significant increase in success probability if one has some \red{\textit{a priori}} information about the input qubit state. In this paper, we \red{analyzed the capabilities of the proposed} linear-optical setup for the \red{state-dependent} qubit amplifier. We determined output state fidelity, gain and success probability \red{as functions} of setup parameters.

\red{Next, we performed a numerical optimization} of success probability depending on target output state fidelity and gain for various input states. This calculation shows that the closer the state is to the pole of Poincaré sphere, the more pronounced is the success probability improvement if fidelity is allowed to drop. Also this effect manifests more strongly in the cases of higher gains.

\red{Furthermore, we performed numerical analysis of success probability as a function of average output state fidelity for several target average gains and levels of \textit{a priori} information about the input state quantified by the Von Mises-Fisher distribution \cite{Fisher}.} The results shows how the \red{maximum} success probability versus fidelity trade-off behaves depending on average gain and \red{\textit{a priori}} information about the input state. To clearly visualize the potential improvement in success probability, we have constructed a specific function of merit that we use to characterize the amplifier in several regimes (various gains and levels of \red{\textit{a priori}} knowledge about the input state). This analysis indicate that success probability can be increased in order of tens of percents depending on the conditions.

\red{Interestingly, we found that in general (for cases other than infinite gain) the success probability of the amplifier does not increase in a monotonic way for decreasing fidelity. This result clearly demonstrates that the success probability of state-dependent amplifiers can be maximally increased without significant drop in output state fidelity. For this reason we believe that our results can stimulate further research on state-dependent qubit amplifiers and their potential applications.}

\begin{acknowledgments}
The authors gratefully acknowledge the support by the Operational Program Research and Development for Innovations -- European Regional Development Fund (project No. CZ.1.05/2.1.00/03.0058). A.~\v{C}. acknowledges project No. P205/12/0382 of Czech Science Foundation. K.~B. and K.~L. acknowledge support by Grant No. DEC-2011/03/B/ST2/01903 of the Polish National Science Centre and K.~B. also by the Operational Program Education for Competitiveness -- European Social Fund project No. CZ.1.07/2.3.00/30.0041 while K.~L. acknowledges the support by Czech Science Foundation (project No. 13-31000P). The authors thank Evan Meyer-Scott, Thomas Jennewein, Norbert L\"{u}tkenhaus, Jan Soubusta and J\'{a}ra Cimrman for inspiration.

\end{acknowledgments}

\end{document}